# Origin of the Enhanced Polarization in La and Mg Co-substituted BiFeO$_3$ Thin Film during the Fatigue Process

Qingqing Ke[1], Amit Kumar[2], Xiaojie Lou[1], Kaiyang Zeng[2] and John Wang[1,a)]

[1] *Department of Materials Science and Engineering, National University of Singapore, Singapore 117574*

[2] *Department of Mechanical Engineering, National University of Singapore, Singapore 117576*

We have studied the polarization fatigue of La and Mg co-substituted BiFeO$_3$ thin film, where a polarization peak is observed during the fatigue process. The origin of such anomalous behavior is analyzed on the basis of the defect evolution using temperature-dependent impedance spectroscopy. It shows that the motion of oxygen vacancies ($V_O^{\bullet\bullet}$) is associated with a lower energy barrier, accompanied by the injection of electrons into the film during the fatigue process. A qualitative model is proposed to explain the fatigue behavior, which involves the modification of the Schottky barrier upon the accumulation of $V_O^{\bullet\bullet}$ at the metal-dielectric interface.

**Keywords**: BiFeO$_3$, oxygen vacancies, polarization fatigue

PACS codes: 77.22.Gm, 77.55.-g, 77.55.Nv

---

a) Author to whom corresponding should be addressed. Electronic mail: msewangj@nus.edu.sg.

BiFeO₃ (BFO) has been extensively studied, as it exhibits both ferroelectric and antiferromagnetic orders at room temperature.[1] The multiferroic behavior makes BFO potentially useful in magnetoelectric and ferroelectric devices, which have a same control knob: polarization switching by an applied electric field.[2] However, the polarization degradation in BFO can result in losing the controllability of magnetic order switching, thus hinders its multifunctional applications. To understand the problem of fatigue failure, several qualitative models have been proposed, including for example seed-inhabitation[3] and wall pinning[4], where the generation and redistribution of charged defects are considered.[5,6] Since oxygen vacancies ($V_O^{\bullet\bullet}$) are the most likely and mobile ionic defects in BFO, the domain wall pinning by $V_O^{\bullet\bullet}$ has been suggested to be a dominant scenario for the polarization degradation.[7]

In contrast to the polarization degradation that has been commonly observed upon repeated switching process, an enhancement in polarization was also reported in certain ferroelectric thin films, for example in heterolayered Pb(ZrTi)O₃/(BiNd)Ti₃O₁₂[8] and Pb(ZrTi)O₃[9] thin films. Sim et al.[8] suggested that it was related to the defects accumulated at the interface between the heterolayers, while Kohli et al.[9] believed that the domain wall de-pinning was responsible for this behavior. Although a similar phenomenon has also been observed in substituted BFO thin film,[10] the exact mechanism is far from being established. In the present work, we have investigated the fatigue behavior of $Bi_{0.9}La_{0.1}Fe_{0.96}Mg_{0.04}O_3$ (BLFM), where an anomalous fatigue behavior is observed. Its polarization is dramatically enhanced at the early stage of repeated switching, in contrast to the commonly expected degradation. By looking into the defect evolution during the ferroelectric polarization switching, we show that the phenomenon is related to the charged defects of $V_O^{\bullet\bullet}$.

The BLFM thin film was deposited by rf magnetron sputtering on Pt-coated Si substrates at 600 °C with a SrRuO$_3$ buffer layer in thickness of 80 nm. The detailed experimental procedure of the film preparation has been reported elsewhere.[11] The resultant BLFM thin film exhibits a thickness of ~300 nm and consists of a single perovskite phase with random-distributed orientation, as verified by X-ray diffraction phase analysis. Prior to electrical measurements, Au dots of 200 μm in diameter were sputtered on the film surface as electrodes. The fatigue behavior was studied by using a Radiant precise workstation (Radiant Technologies, Medina, NY) at varying temperatures. Temperature-dependent impedance spectroscopy was conducted using a Solartron impedance analyzer, where the measurement was made from room temperature up to 180 °C over the frequency range of 0.1–10$^6$ Hz. To investigate the local surface charge conditions, Kelvin probe force microscopy (KPFM) was performed under ambient condition using an atomic force microscopy (MFP-3D, Asylum Research, USA) with a platinum coated silicon cantilever (radius of 15 nm with a spring constant of 2 N/m and a resonant frequency of 70 kHz). KPFM images over a scan area of $2 \times 1 \mu m^2$ were acquired under an ac voltage of 3 V at the lift height of 40 nm in trapping mode.

Fig. 1 shows the polarization switching under a bipolar triangle pulse with amplitude of 400 kV/cm at 100 kHz, where an unusual fatigue behavior is observed in the BLFM thin film. The polarization remains stable up to 10$^5$ cycles at each temperature. However, it is dramatically enhanced with number of switching cycles ($N$) at >10$^5$, increasing to approximately four times of the initial value upon the application of ~10$^8$ switching cycles at 60 °C. A polarization peak is observed at each temperature with a shift towards smaller $N$ and the peak becomes stronger as temperature increases. Note that the film exhibits a random distributed orientation. The enhanced polarization (e.g. the value of ~120 μC/cm$^2$ measured for the film fatigued at 60 °C, which is higher than that measured for the epitaxial (111)-oriented BFO films, ~100 μC/cm$^2$ [12]) can not be simply ascribed to the ferroelectric response.

Fig. 2 shows the dielectric constant as a function of frequency for the BLFM thin film measured at different stages of polarization switching. The fresh sample shows a rather stable value against frequency. With increasing the number of switching cycles, the dielectric constant of the fatigued sample is dramatically enhanced at the low frequency regime from 1 to 1000 Hz, which is caused by the dielectric response of space charges to the applied electric field. The variation in dielectric spectra clearly indicates an evolution of charged defects at the different fatigue stages in the BLFM thin film. The displacement of charged defects, which involves energy dissipation, can give rise to a hysteresis behavior and contribute to polarization.[13] Consequently, the anomalous fatigue behavior observed for BLFM in the present work shall be closely correlated to the variation of defects during the fatigue process.

Any motion of charged defects inevitably interacts with their neighbors, leading to a relaxation process upon the application of an ac field. We have therefore conducted a contrast study on the relaxation behavior between the fresh and the fatigued samples in order to reveal some details about the evolution of defects during the fatigue process. The frequency-dependent imaginary part of modulus ($M''$) for the fresh sample is shown in Fig. 3(a), where a single relaxation peak is observed at each temperature and shifts towards higher frequencies with increasing temperature, indicating a thermally-activated process. To determine the nature of defects involved in the fresh sample, the activation energy can be obtained using the Arrhenius law [$f = f_0 \exp(-\frac{E_a}{K_B T})$], where $f$ is the hopping frequency, $f_0$ is the pre-exponential factor, and $E_a$ is the activation energy. As shown in the fitting results plotted in Fig. 3(b), the activation energy thus obtained is 0.9 eV, which is comparable to 0.99 eV measured for Bi:SrTiO$_3$,[14] indicating that $V_O^{\bullet\bullet}$ are the main defects existing in the fresh sample. The formation process of $V_O^{\bullet\bullet}$ can be described using the Kröger–Vink notation as:[11]

$$2MgO \rightarrow 2Mg'_{Fe} + 2O^\times + V_O^{\bullet\bullet} \quad (1)$$

where $V_O^{\bullet\bullet}$ represent the second-ionized oxygen vacancies.

In contrast to the fresh sample, the fatigued film shows some new features in $M''$ spectra as plotted in Fig. 3(c). At low temperatures ranging from 20 to 100 °C, two $M''$ peaks can be clearly observed, with one being located in the low frequency region (below 10 Hz) which is labeled as peak I, and the other one in the high frequency region (from 10-1000 Hz) labeled as peak II. Moreover, peak I disperses much more strongly and approaches to peak II with increasing temperature, as indicative of two different thermal activation processes. When temperature is raised to above 120 °C, the two peaks merge into one stronger peak (labeled as peak III). The occurrence of these three peaks suggests different relaxation processes occurring in the fatigued sample in association with defect variation in the different temperature range. Their characteristics can be identified by Arrhenius plot as shown in Fig. 3(d). The activation energies for these three relaxation processes are calculated to be 0.49, 0.24, and 0.7 eV for peak I, II, and III, respectively.

An activation energy of 0.47 eV has been reported for $Bi_{0.9}La_{0.1}Fe_{0.95}Ti_{0.05}O_3$ (BLFT) thin film, which is ascribed to the movement of oxygen vacancies in the single-ionized state ($V_O^{\bullet}$).[15] Considering the similarity in composition and the comparable value of activation energy measured between BFLM and BLFT, one can conclude that the relaxation process for peak I shares the same mechanism as that in BLFT. For peak II, the activation energy of 0.24 eV is close to the energy (~0.1 eV) for the first-ionization process of oxygen vacancies,[16] suggesting that in this temperature range free electrons are generated from the first-ionization of $V_O$, which can be expressed as:

$$V_O \rightarrow V_O^{\bullet} + e' \qquad (2)$$

where $V_O$ is the oxygen vacancy in the neutralized state. With increasing temperature, the electrons trapped by $V_O^{\bullet}$ will be further released in the second-ionization process[11]

$$V_O^{\bullet} \rightarrow V_O^{\bullet\bullet} + e' \qquad (3)$$

As has been reported by Long and Blumenthal,[17] the second-ionization energy of $V_O$ is $E_d$= ~1.4 eV. Therefore the activation energy for electrons released from $V_O^{\bullet}$ should be 0.7 eV ($E_d$/2), which is consistent with the value measured for peak III in the present work. Moreover, the activation energy of 0.7 eV also falls in the range from 0.6 to 1.2 eV that was ascribed to the movement of $V_O^{\bullet\bullet}$.[18] Therefore, two possible mechanisms corresponding to the peak III may act in parallel: (i) electrons released from $V_O^{\bullet}$, and (ii) movement of $V_O^{\bullet\bullet}$ responding to the external field. We note that the activation energy measured for the displacement of $V_O^{\bullet\bullet}$ in the fatigued sample is much lower than that in the fresh one. A likely reason is that an increase number of $V_O^{\bullet\bullet}$ is introduced into the film as a result of the lattice distortion during the fatigue process,[3] leading to a stronger correlation among the defects and prompting the motion of $V_O^{\bullet\bullet}$ with a lower energy barrier.

From the ionization process of $V_O$ (shown in Eq. 2 and Eq. 3) mentioned above, one can trace the neutralization of $V_O^{\bullet\bullet}$ during the fatigue process, which is further confirmed by KPFM study in the present work. To exclude the contribution of the bound charges from domains, a small amplitude of 80 kV/cm (much lower than the coercive field of 330 kV/cm as indicated in Fig. 4(a)) was applied to pole the sample surface at a frequency of 100 kHz for 1 minute. The white image in Fig. 4(c) was thus observed compared to the brown-colored-unbiased surface shown in Fig. 4(b), indicating a much higher surface potential in the poled sample. The surface potential arises as a result of the compensation between internal charges and surface screen charges[19], when the polarization bound charges are not considered. As certain negative screen charges exist on the as-deposited ferroelectric film surface,[19] one can therefore conclude that the enhancement in surface potential observed in the poled sample is due to the positive space charges (e.g., the involvement of $V_O^{\bullet\bullet}$) accumulated at the interface, which could well compensate for the negative surface charges as illustrated in Fig. 4(d). In the prolonged fatigue process, as more and more $V_O^{\bullet\bullet}$ accumulate at the interface, the width of the Schottky barrier decreases[20] (i.e., the Debye length decreases as $n^{-1/2}$, where $n$ is the density of accumulated charges). It enhances the possibility of electron injection from the electrode, thus leading to the neutralization of $V_O^{\bullet\bullet}$ at the interfaces.

Our central idea is that the migration and neutralization of $V_O^{\bullet\bullet}$ are responsible for the anomalous fatigue behavior observed for BLFM. Upon the application of an ac electric field (i.e., amplitude of 400 kV/cm), the defects pair between $V_O^{\bullet\bullet}$ and $Mg_{Fe}^{'}$ can be dissociated, and $V_O^{\bullet\bullet}$ as mobile defects will migrate towards the cathode, where they pile up. This is equivalent to a macroscopic dipole with a length of the film thickness and thus makes contribution to the total polarization. With increasing the number of switching cycles, more and more $V_O^{\bullet\bullet}$ oscillate with the ac field, and the polarization is continuously enhanced. However, once the concentration of $V_O^{\bullet\bullet}$ within the interfacial layer reaches a critical value, the width of the Schottky barrier will become thin enough allowing electrons to surge into film and thus neutralize $V_O^{\bullet\bullet}$, giving rise to a reduction in polarization for the fatigued thin film. At elevated temperatures, the motion of $V_O^{\bullet\bullet}$ will be further facilitated with a higher mobility. As a result, it is much easier for $V_O^{\bullet\bullet}$ to drift towards the electrode, where they accumulate, further decreasing the barrier width and leading to a shift of the polarization peaks towards smaller numbers of switching cycles as shown in Fig. 1.

To summarize, an anomalous fatigue behavior is observed in the BLFM thin film, where the polarization is enhanced rather than degraded upon the repeated switching cycles. The phenomenon is investigated by looking into the evolution of charged defects in the film during the fatigue process, where the mobility of $V_O^{\bullet\bullet}$ is evolved with a change in energy barrier and electrons are injected into film by the repeated switching cycles.

The motion of $V_O^{\bullet\bullet}$ under an applied field gives rise to an initial increase in polarization. The subsequent injection of electrons due to the modification of Schottky barrier however neutralizes and decreases the concentration of $V_O^{\bullet\bullet}$, resulting in a reduction in the polarization upon the prolonged fatigue.

Captions for Figures:

FIG. 1. (Color online) Fatigue behavior of BLFM thin film measured at various temperatures.

FIG. 2. (Color online) Dielectric constant as a function of frequency for the BLFM thin film that was subjected to different numbers of switching cycles.

FIG.3. (Color online) Variation of $M''$ as a function of frequency, for (a) the fresh sample, (c) the fatigued sample (after $10^9$ switching cycles). The corresponding Arrhenius plots of the hopping frequency against temperature are shown in (b) and (d), respectively.

FIG. 4. (Color online) (a) Room temperature hysteresis loops of the fresh BLFM thin film under various electric fields, where the inset shows the remanent polarization and coercive field as a function of electric field. KPFM surface potential distribution of (b) the poled sample, (c) the as-deposited sample. (d) Schematic depiction of the charges distribution after an ac poling process. The red and blue circular charges represent the surface screen charges and positive charges inside the film. The scale bar represents 0.5 μm.

FIG. 1

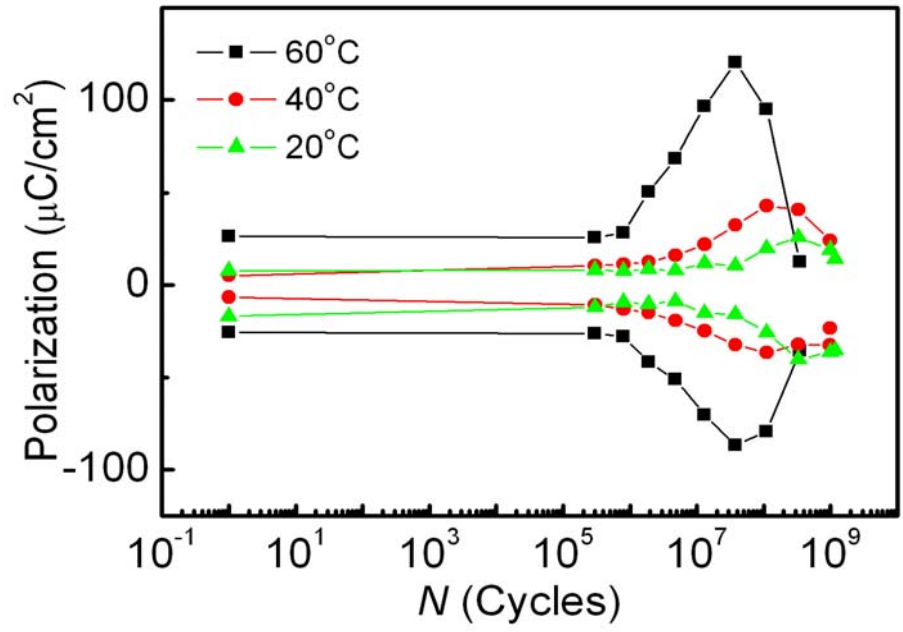

FIG. 2.

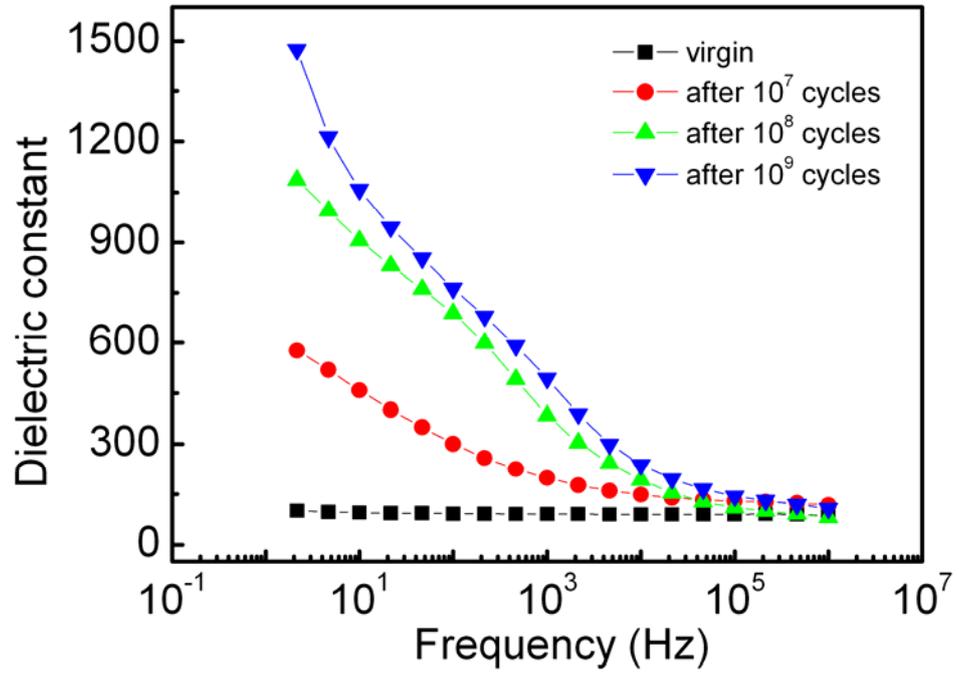

FIG. 3

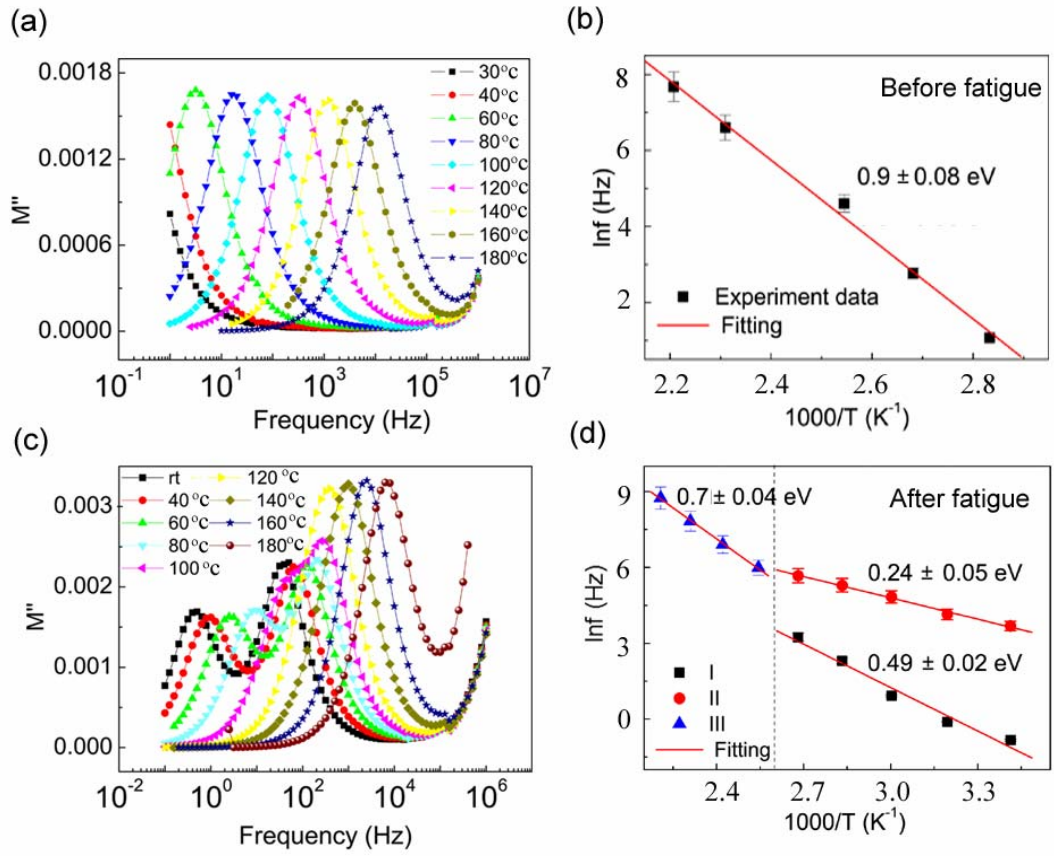

FIG. 4

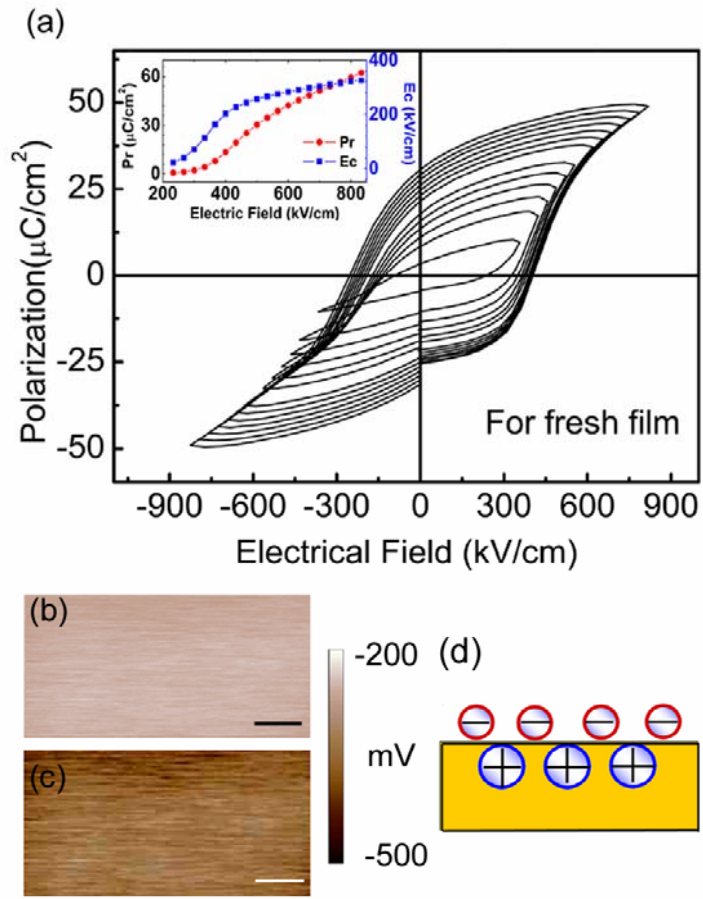